# Can Twitter Predict Royal Baby's Name ?

Bohdan Pavlyshenko

*Ivan Franko Lviv National University,Ukraine, b.pavlyshenko@gmail.com*

**Summary**

In this paper, we analyze the existence of possible correlation between public opinion of twitter users and the decision-making of persons who are influential in the society. We carry out this analysis on the example of the discussion of probable name of the British crown baby, born in July, 2013. In our study, we use the methods of quantitative processing of natural language, the theory of frequent sets, the algorithms of visual displaying of users' communities. We also analyzed the time dynamics of keyword frequencies. The analysis showed that the main predictable name was dominating in the spectrum of names before the official announcement. Using the theories of frequent sets, we showed that the full name consisting of three component names was the part of top 5 by the value of support. It was revealed that the structure of dynamically formed users' communities participating in the discussion is determined by only a few leaders who influence significantly the viewpoints of other users.

**Introduction**

Twitter messages are characterized by high density of contextually meaningful keywords. This feature conditions the availability of the study of microblogs by using data mining in order to detect semantic relationships between the main concepts and discussion subjects in microblogs. Very promising is the analysis of predictive ability of time dependences of key quantitative characteristics of thematic concepts in the messages of twitter microblogs.

The peculiarities of social networks and users' behavior are researched in many studies. In [1,2], the microblogging phenomena were investigated. In [3], it is shown that social networks differ structurally from other types of networks. Users' influence in twitter was studied in [4]. Users' behavior in social networks is analyzed in [5]. In [6], the methods of opinion mining of twitter corpus were analyzed. Several papers are devoted to the analysis of possible forecasting of events by analyzing messages in microblogs. In [7], it was studied whether public mood as measured from large-scale collection of tweets posted on twitter.com, is correlated or predictive for stock markets. In [8], it is shown that a simple model built from the rate, at which tweets are created about particular topics, can outperform market-based predictors. In [9], films sales based on the discussions in microblogs are analyzed. In [10], the twitter activity during media events was investigated.

One of events of summer 2013 under most discussion was the birth of British crown prince. There were lots of discussions of this event in social networks, in particular the debates as to the prince's possible name. From the position of social networks, it is interesting to what extent such discussions can be predictable. Is it possible to predict the crown prince's name on the basis of the analysis of tweets?

It is really not a serious problem, if to take it literally. But the main goal of this study is to test whether there is any correlation between social network users' opinions and the decisions that can be made by individuals who are highly influential in certain spheres of the society.

In this paper, we will study the existence of possible correlation between twitter users' social meaning and the decision-making of the persons who are highly influential in the society. We will conduct the analysis on the example of the discussion of probable name of the British crown baby, born in July, 2013. In this analysis, we will use the methods of quantitative processing of natural language and the theories of frequent sets. We will study the time dynamics of keyword frequencies and the structure of users' communities, formed dynamically during the discussion.

**Theoretical Model**

Let us consider a model that describes microblogs messages. We have chosen some set of keywords which specify the themes of messages and are present in all messages. It may be, for example, the tags '#RoyalBaby', '#RoyalBabyWatch', "#Royals", etc.
Then we define a set of microblogs messages for the analysis:

$$TW^{kw} = \{tw^{(kw)}{}_i \mid kw_j \in tw_i, \ kw_j \in Keywords\}. \quad (1)$$

The keywords define the subject matter of analyzed tweets array. As a quantitative characteristics for name analysis, we use the frequency of the tweets which contain the mane under analysis:

$$F_{name}(Name) = \frac{|\{tw^{(kw)} \mid Name \in tw^{(kw)}, time_s^{tw} \in [time_1^{tw}, time_2^{tw}]\}|}{|\{tw^{(kw)} \mid time_s^{tw} \in [time_1^{tw}, time_2^{tw}]\}|} \quad (2)$$

where $time_s^{tw}$ is the time of tweet sending, $[time_1^{tw}, time_2^{tw}]$ is the time window of the analysis of tweets containing the keyword.

Our next step is to consider the basic elements of the theory of frequent sets. Each tweet will be considered as a basket of key terms

$$tw_i = \{w_{ij}^{tw}\}. \quad (3)$$

Such a set is called a transaction. We label some set of terms as

$$F = \{w_j\}. \quad (4)$$

The set of tweets, which includes the set $F$ looks like

$$TW_F^{kw} = \{tw_r \mid F \in tw_r; r = 1,...m\}. \quad (5)$$

The ratio of the number of transactions, which include the set $F$, to the total number of transactions is called a support of $F$ basket and it is marked as $Supp(F)$:

$$Supp(F) = \frac{|TW_F^{kw}|}{|TW^{tw}|}. \quad (6)$$

A set is called as frequent, if its support value is more than the minimum support that is specified by a user

$$Supp(F) > Supp_{min}. \quad (7)$$

Given the condition (7), we find a set of frequent sets

$$L = \{F_j \mid Supp(F_j) > Supp_{min}\}. \quad (8)$$

For identifying frequent sets, an Apriori algorithm [11, 12] is mainly used. It is based on the principle that the support of some frequent set does not exceed the support of any of its subsets.

**Experimental part**

Here we are considering the sequence of conducted analysis. We have been loading the tweets in the time period from July 19 to July 25, 2013. To single out the tweets concerning the analyzed subject of Royal crown prince's name, we used such keywords as: '#RoyalBaby', '#RoyalBabyWatch', "#Royals", '#RoyalBabyName', "#goodluckKate", "Kate Middleton". For downloading the tweets, we used Twitter API and Python package python-twitter. We recorded downloaded data into SQLite database. The analysis was conducted in the environment of statistical R calculations with the use of additional packages, in particular: xts, tm, arules, "arulesViz", RSQLite, igraph. On the basis of obtained data, we singled out the array of tweets with the keyword "name". Then we formed a frequency dictionary and deleted all stop-words out of it. While analyzing the frequency dictionary, we singled out the words representing names. A set of potentially possible names looks like the following:

Snames ={"Albert", "Alexander", "Andrew", "Arthur", "Boris", "Charles", "Edward", "Freddy", "George", "Harry", "Henry", "James", "Joffrey", "John", "Joseph", "Louis", "Michael", "Philip", "Richard", "Rudiger", "Spencer", "Stuart" }

(9)

We already know that the Royal baby's name is Prince George of Cambridge. The Crown Prince's full name is George Alexander Louis. So, let us consider the time dynamics of the name George. As a quantitative characteristic, we examine the frequency of tweets with given key name, which we calculate by the formula (2). Figure 1 shows the time dynamics of the frequency of tweets containing the name George.

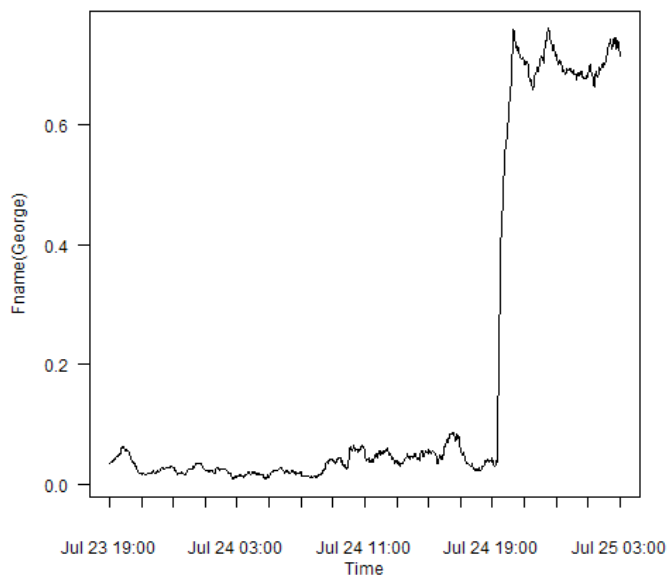

*Fig.1 The time dynamics of the frequency of tweets containing the name George.*

Time is given in the standard of Coordinated Universal Time (UTC). In this figure, we can see a sharp jump which apparently corresponds to the time of the announcement of newborn prince's name. The time jump occurs in the time period from 19:20 to 19:30 on July 24, 2013. Let us consider the array of tweets, which corresponds to the time period before this abrupt jump. For the analysis, we select the tweets that were sent before 18:00 on July 24, when the prince's name was still unknown. We examined 22411 tweets which were downloaded before the mentioned jump starting from July 19 and which contained at least one of above mentioned names and also the keyword 'name'. Figure 2 shows the frequency of tweets containing the names under analysis which are included into a set of names Sname (9) in the period before the official announcement of prince's name. The names are listed in descending order of Fname frequency. As the obtained diagram shows it is the name George that is leading.

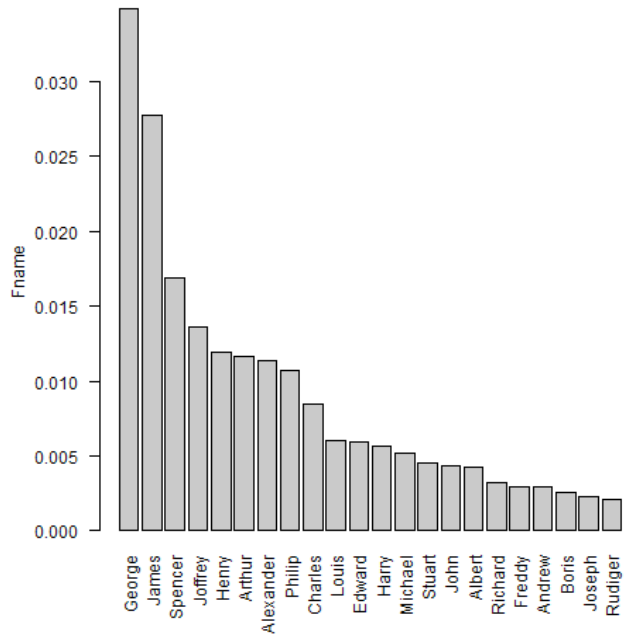

*Fig.2 The distribution of the names in tweets in the period before the prince's name announcement.*

This indicates that by analyzing tweets, one could predict the principal name of the Crown prince. It means that there is a strong correlation between bloggers' viewpoints in the discussions of prince's name and the real decision-making of the Royal family as to this very subject matter. The next two prince's names Alexander and Louis are situated on the seventh and tenth positions in the frequency row of names, respectively. Figure 3 shows the time dynamics of the second name in the rate - James in the frequency row of names (Fig. 2). As we can see, at the very moment of the official announcement of the name there is sharp decline of the frequency of tweets with the name James.

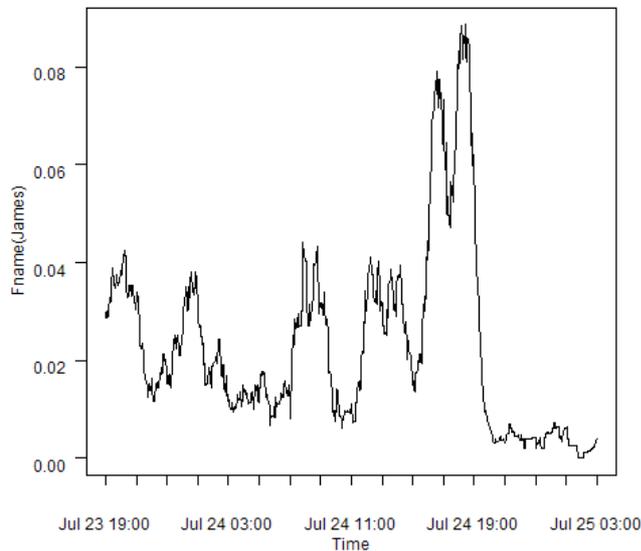

*Fig.3 Time dynamics of the frequency of tweets with the name James.*

The same decline is also observed for other names that are not the part of the full name of prince George Alexander Louis. Figure 4 shows the distribution of names in tweets after the prince's name was announced. We took into consideration 11620 tweets that were sent after 21:00 on July 24 and contained the keyword "name". There is an expected result, the first three names in the list correspond to the component names of the prince.

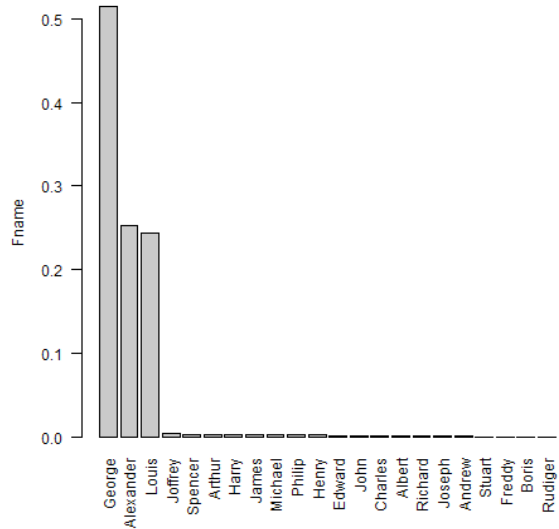

*Fig.4 The distribution of names in tweets after the prince's name was announced.*

Let's go back to the analysis of tweet array downloaded before the name the announcement and analyse whether it was possible to predict the full name that consists of three components. Let us consider the names which occured in tweets simultaneously. For our analysis, we use the elements of the theory of frequent sets. Have a look at those tweets in the analyzed array, which contain the frequent sets of three keywords belonging to the set of analyzed names. Using the algorithm a priori [11,12] we found the frequent sets of words that represent potential names in the array of tweets. The value of support of the frequent sets, we calculated by the formula (6). The frequent sets with the highest support value are shown in the table 1:

Table 1 Frequent itemsets with the highest support values

| # | item | support |
|---|---|---|
| 1 | alexander, george, james | 0.120646766 |
| 2 | george, henry, james | 0.106965174 |
| 3 | george, james, louis | 0.089552239 |
| 4 | alexander, james, louis | 0.084577114 |
| 5 | alexander, george, louis | 0.084577114 |
| 6 | george, henry, louis | 0.079601990 |
| 7 | alexander, henry, james | 0.077114428 |
| 8 | alexander, george, henry | 0.077114428 |
| 9 | alexander, henry, louis | 0.074626866 |
| 10 | henry, james, louis | 0.074626866 |
| 11 | arthur, george, james | 0.018656716 |
| 12 | charles, james, spencer | 0.016169154 |
| 13 | george, james, spencer | 0.014925373 |
| 14 | charles, george, philip | 0.013681592 |
| 15 | george, james, richard | 0.012437811 |

As the obtained data show, the frequent set of words which contains all three names of the prince (George, Alexander, Louis) is the part of first five frequent sets, ordered by the support value. Speaking about two prince's component names, they are in the top three of the frequent set list. Figure 5 shows the graph of the formation of the first ten frequent sets by the support value. Out of five names being the parts of first ten frequent sets, three names are the component names of the prince.

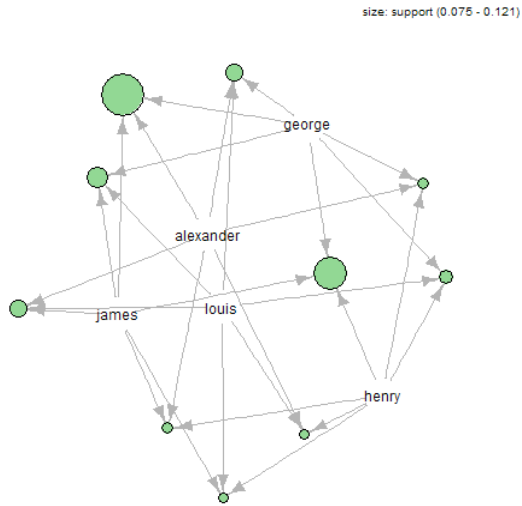

*Fig.5 The graph of the formation of first 10 frequent sets with the biggest support value.*

As we can see from the data, the theory of frequent sets allows to get a more accurate prediction for the full name as opposed to the received frequency list of names which makes it possible to predict a major name only.

Consider the set structure of the users who participated in the discussion of the prince's name. To identify the communities that were formed dynamically in the discussion under analysis, we used a fast greedy modularity optimization algorithm, described in [13]. To build a graph, we used a Fruchterman-Reingold algorithm[14]. This algorithm belongs to force algorithms, or spring algorithms. The character of the graph is due to the model which is used in force algorithms. The distinctive feature of the model is that its vertices are considered as the balls, affected by repulsive forces; and the edges are considered as spring models that attract the vertices which are connected by these edges . In the tweet arrays, we have found 6919 users that sent 37191 tweets. These tweets mentioned 2645 users. An essential part of these mentions is relates to retweets. For further analysis, we take active users who sent more than on tweet in the process of discussion or who were mentioned in tweets more than once. We have found 2,300 active users who sent more than one tweet, and 923 users who were mentioned in tweets more than once. Figure 6 shows the graph of users' interrelations, the shades of colors on it mark the users' communities. On this graph, we can see that there are several numerous users' communities.

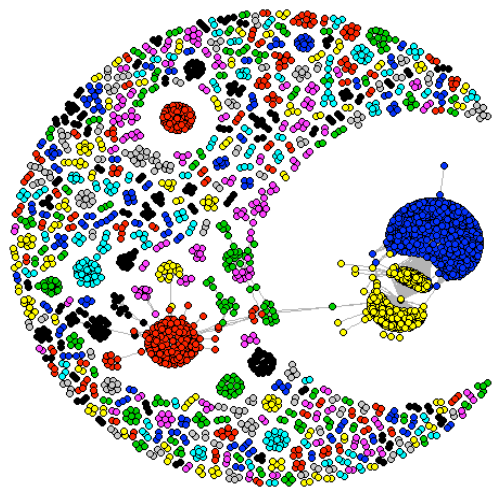

*Fig.6 Revealed users' communities.*

Our next step is to conduct the analysis after removing the most popular users that were mentioned in tweets 100 times or more. We have found only 6 such users. Having removed these users from the analysis, we received the community graph, which is shown on Figure 8. Removed users constitute nearly 0.2% of all the users mentioned in tweets. As follows from the obtained data, that if to remove only the most popular users from the analysis, the community structure will be changed significantly, and only numerous small communities will be left. We also detected the users who were mentioned 50 or more times, there are 16 such users. Having removed these users from the analysis, we received the graph with the identified communities, it is shown on Fig.9

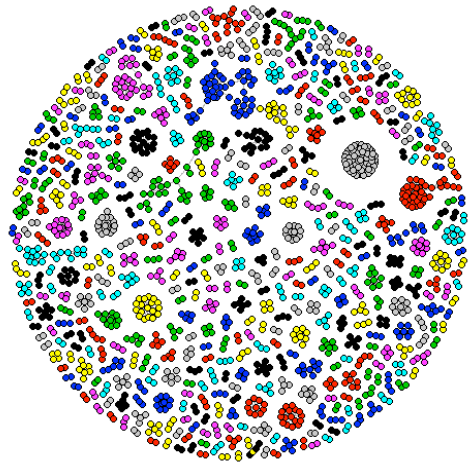

*Fig.7 Users' communities without six most popular users.*

**Conclusions**

The results of the study demonstrate that tweets mining could predict the Royal baby's name. We showed that the major name of newborn Prince George was dominant in the spectrum of names before the official announcement. It follows from the obtained data that the theory of frequent sets allows to get a more precise prediction for the full name if to compare with the analysis of the name frequency range which allows to predict a major name only. The three prince's component names George, Alexander, Louis form a frequent itemset of words and this itemset was the part of the top 5 largest frequent itemsets by the support value. We also showed that the structure of dynamically formed users' communities that participated in the discussion is defined by only several leaders who have a significant influence on the position of other users. What do these results mean? It is really not a serious problem, if to take it literally. But the main goal of this study is to test whether there is any correlation between social network users' opinions and the decisions that can be made by individuals who are highly influential in certain spheres of the society. In our studies, we revealed that such a correlation does exist. This means that there is a certain correlation between the bloggers' viewpoints and the decision-making of the Royal family as to the prince's name.